\pdfoutput=1 
\documentclass[aps,prb,twocolumn,groupedaddress,amsmath]{revtex4}

\usepackage{graphicx}
\usepackage{amssymb}
\usepackage{bm}
\usepackage{amsmath,amsfonts,latexsym}
\usepackage{comment}

\usepackage{bm} 

\newcommand*{\br}{\mathbf{r}}

\newcommand*{\cN}{{\cal N}}
\newcommand*{\cL}{{\cal L}}
\newcommand*{\cH}{{\cal H}}
\newcommand*{\cT}{{\cal T}}

\newcommand*{\phop}{\phi^{\vphantom{\dagger}}}
\newcommand*{\phdop}{\phi^\dagger}

\DeclareMathOperator{\tr}{tr}
\newcommand{\vect}[1]{\bm{#1}}

\begin{document}

\title{Low energy dynamics of spinor condensates}

\author{Austen Lamacraft} 
\affiliation{Department of Physics, University of Virginia,
Charlottesville, VA 22904-4714 USA}
\date{\today}
\email{austen@virginia.edu}
 
\date{\today}

\begin{abstract}

We present a derivation of the low energy Lagrangian governing the dynamics of the spin degrees of freedom in a spinor Bose condensate, for any phase in which the average magnetization vanishes. This includes all phases found within mean-field treatments except for the ferromagnet, for which the low energy dynamics has been discussed previously. The Lagrangian takes the form of a sigma model for the rotation matrix describing the local orientation of the spin state of the gas.

\end{abstract}

\maketitle


\section{Introduction}

\subsection{Spin ordering in ultracold gases}

In recent years the field of ultracold atomic physics has attracted the attention of a great many condensed matter theorists, largely due to the prospect of finding novel realizations of many-body systems. Part of the appeal doubtless lies in the exquisite experimental control that may be exercised over the parameters of a system in which many of the complicating factors familiar from the solid state (disorder, phonons, etc.) are absent. Certain \emph{intrinsic} aspects of ultracold atomic gases -- not dependent on the specifics of the experimental setting -- are qualitatively novel, however, and without antecedent in the study of condensed matter. In this latter category we may place the possibility of spontaneous ordering of the spin degrees of freedom in a Bose gas. Indeed, prior to the `ultracold revolution' the only Bose superfluid that could be studied in the laboratory was $^4$He, which has zero spin. With the advent of optical trapping of Bose condensates of alkali atoms, which allows for a fully rotationally invariant setting, the experimental study of spin ordering within a hyperfine multiplet came within reach~\cite{Stenger1998,ketterle2000}. 

The earliest theoretical works motivated by these developments explored possible ordered phases using a mean field description, in both the spin 1 and spin 2 hyperfine multiplets~\cite{ho1998,ohmi1998,ciobanu2000,ueda2002}. In this description the concept of spontaneous symmetry breaking plays a central role. Up until very recently, however, there were no experiments in which this spontaneous ordering was apparent. The reason for this is that the simplest experimental protocol for the investigation of the hyperfine state of the gas is to apply a magnetic field gradient to split a gas cloud into different components in a Stern-Gerlach experiment~\cite{Stenger1998,ketterle2000}. The different components are subsequently imaged to determine their (relative) occupancies. 
This technique naturally imposes a quantization axis, and any information concerning the \emph{coherence} between different hyperfine levels is lost. Thus magnetic alignment in the plane perpendicular to this axis, which depends on the relative phase of these different levels, cannot be observed. 



%
The characterization of magnetic ordering in atomic gases has taken a leap forward in the last few years thanks to the work of the Berkeley group, who demonstrated \emph{in situ} dispersive imaging of the transverse magnetization of a gas of $^{87}$Rb in the spin 1 multiplet~\cite{higbie2005}, and subsequently employed this technique to investigate a number of fascinating aspects of this system, including the dynamics of spontanteous symmetry breaking, defect production, and the role played by magnetic dipole forces~\cite{sadler2006,Vengalattore2008,Vengalattore2009}.

The above developments illustrate two important needs. Firstly, imaging of the spin order was necessary to bring much of this new physics out into the open. Secondly, the nonequilibrium character of most experiments requires that the mean field theory of equilibrium ordered states be supplemented with a \emph{dynamical} description of the relevant order parameters. 

It is our hope that the next few years will see the development of imaging techniques capable of detecting some of the spin orders to be discussed in Section~\ref{sec:spin}, of which an average magnetization is only the simplest. The aim of this work is to address the second need: uncovering the low energy description of the order parameter. For the case of the Bose ferromagnet, which is appropriate to the spin 1 $^{87}$Rb system, this description was provided in an earlier paper~\cite{Lamacraft2008}. In this work we will focus instead on states with vanishing average magnetization. For reasons that will be become clear in the following sections these cases are qualitatively different.

\subsection{Low energy descriptions}

In an ordered phase of matter we expect that the low energy degrees of freedom consist of variations of the order parameter on some manifold of symmetry broken states (Goldstone modes), together with any conserved quantities. In our earlier work on the Bose ferromagnet~\cite{Lamacraft2008} the degrees of freedom were the local magnetic moment $\vect{m}$ and the superfluid velocity $\vect{v}$. In the long wavelength limit these were found to obey the coupled equations
\begin{equation}\label{LLE}
\frac{D\vect{m}}{Dt}-\frac{\hbar^2}{2m}\vect{m}\times \nabla^2\vect{m}=0
\end{equation}
\begin{eqnarray}\label{v_fix}
\nabla\cdot \vect{v}=0,  \qquad
\nabla\times\vect{v}=\frac{\hbar s}{2m} \epsilon_{abc} m_a \nabla m_b \times \nabla m_c
\end{eqnarray}
where $\frac{D}{Dt}=\frac{\partial}{\partial t}+\vect{v}\cdot\nabla$ is the convective derivative. Eq.~(\ref{LLE}) is a modified Landau-Lifshitz equation and gives rise to quadratically dispersing spinwaves when linearized on around a solution $\vect{m}=\mathrm{const.}$.

The quadratic dispersion is a consequence of the two transverse deviations of the order parameter being canonically conjugate. For the phases with vanishing average magnetization that are the focus of this work, the conjugate variables involve deviations from the order parameter manifold. This results in linearly dispersing Goldstone modes. The situation is analogous to the case of spin waves in an antiferromagnet, where the conjugate variables are the difference in magnetization on neighboring sites -- the N\'eel order parameter -- and the sum.

It also follows from the vanishing of the average magnetization that the order parameter dynamics and superfluid flow are decoupled except for a global topological constraint. Again, this is quite different from the case of the ferromagnet, where the two are coupled together in the equations of motion. It follows that we can write a Lagrangian for the spin degrees of freedom only. This Lagrangian is expressed in terms of a rotation matrix $R$ that specifies the local orientation of the spin state relative to some reference state. By expressing the matrix elements of $R$ in terms of an orthonormal triad $R_{ab}=\left(\vect{e}_b\right)_a$, with $\vect{e}_a\cdot \vect{e}_b=\delta_{ab}$, the spin Lagrangian may be written
\begin{equation}\label{L_genspin}
\cL_{\mathrm{spin}}=\frac{1}{2}\sum_{a=1}^3\left[\tilde I_a\left(\partial_t \vect{e}_a\right)^2-\tilde g_a\left(\nabla \vect{e}_a\right)^2\right]
\end{equation}
Here $\tilde I_a$ are $\tilde g_a$ are some constants to be specified later, which depend on the ordered phase in question. The Lagrangian Eq.~(\ref{L_genspin}) is the main result of this paper. The only other phase for which the spin Lagrangian was previously obtained is the polar phase of the spin 1 gas, to be discussed below, in which $\tilde I_a=\tilde g_a=0$, $a=2,3$, giving the familiar $O(3)$ model~\cite{zhou2001}.

\subsection{Outline of this paper}

The structure of the remainder of this paper is as follows. In Section~\ref{sec:basics} we review the basic description of a spinor condensate, and in particular the structure of the interaction Hamiltonian. Section~\ref{sec:spin} describes in detail the ground state manifolds of the different phases of spinor condensates, beginning with the simplest non-trivial case, spin 1, before introducing the Majorana (or stellar) representation that is very useful in visualizing spin states. The global structure and local geometry of the order parameter manifolds are then introduced, as well as a parametrization for the  dynamical variables conjugate to the order parameter. After this the derivation of the low energy Lagrangian is a relatively simple matter, and is described in Section~\ref{sec:low} along with the derivation of the equations of motion. We have tried to keep the presentation pedagogical throughout, though at various points there are parenthetical technical comments that the casual reader should feel free to ignore.

In a related work, Barnett \emph{et al.}~\cite{barnett2009} have derived the full equations of motion of a spinor condensate in terms of the Majorana representation and applied a group-theoretical analysis to the determination of all normal modes about an ordered ground state. We pursue a complimentary goal of obtaining the full \emph{nonlinear} Lagrangian for the Goldstone modes only.

\section{Basics of spinor Bose condensates}\label{sec:basics}

\subsection{Lagrangian}

One could take the point of view that the description of the dynamics of a dilute spinor gas is no different from its spinless counterpart, being governed by the time-dependent Gross-Pitaevskii equation, with Lagrangian density
\begin{equation}\label{GP_action}
\cL=i\phdop\partial_t\phop-\cH(\phdop,\phop)
\end{equation}
(we set $\hbar=m=1$ from now on). For a spin $s$ gas $\phop$ is a $2s+1$ component spinor and the Hamiltonian density has the form
\begin{equation}\label{H_dens}
\cH(\phdop,\phop)=\frac{1}{2}\nabla\phdop\nabla\phop+\cH_{\mathrm{int}}(\phdop,\phop),
\end{equation}
where the first term is the kinetic energy. The interaction part $\cH_{\mathrm{int}}(\phdop,\phop)$ is quartic in $\phop$ and its form will be given below for $s=1$ and $2$. We will not discuss the influence of the trapping potential, save to assume that it preserves rotational symmetry. For stationary solutions of the form $\phop(\br,t)=e^{-i\mu t}\phop(\br)$ the time-dependent description reduces to the time-independent Gross-Pitaevskii theory, with $\mu$ the chemical potential. Instead of treating the action $S=\int d\br dt\, \cL$ classically, we can interpret it as the quantum action in a coherent  state path integral. Little that we will have to say will depend upon this distinction.

In fact this superficial similarity between the spinless and spinful problems is quite misleading. The ground state in a uniform system corresponds to some constant $\phop$. In the spinless case this fixes $\phop$ up to a phase once the density $\rho=\phdop\phop$ is specified. But in the spinful case we still have to find the correct `direction' of $\phop$ in the complex $2s+1$ dimensional spinor space, determined by minimizing the interaction Hamiltonian $\cH_{\mathrm{int}}(\phdop,\phop)$. The interactions will be assumed to respect rotational symmetry, so this minimum is only unique up to rotation (specified by the three Euler angles, say). Choosing this rotation -- starting from some arbitrary reference state -- specifies the spontaneous breaking of rotational symmetry in the ground state. The slow variation of this rotation in space and time constitutes the low energy dynamics of the system, the description of which is the subject of this work. To characterize these low energy manifolds we must first specify the form of $\cH_{\mathrm{int}}$.

\subsection{Interaction Hamiltonian}\label{sec:Hint}

The structure of $\cH_{\mathrm{int}}$ has been discussed in many works, starting with the first papers treating spinor Bose condensates. We will therefore keep the following discussion relatively brief. Low energy scattering between a pair of bosons occurs in the $s$-wave channel only, and can be treated as a $\delta$-function interaction, characterized by a set of interaction constants $g_S$, $S=0,2,\ldots 2s$ for a pair of bosons with total spin $S$. Bose symmetry dictates that the interaction vanishes for odd total spin.

For spin 1, the resulting interaction may be presented in the form~\cite{ho1998,ohmi1998}
\begin{equation}\label{spin1Hint}
\cH_{\mathrm{int}}=\frac{c_0}{2}\left(\phdop\phop\right)^2+\frac{c_2}{2}\left(\phdop\vect{S}^{(1)}\phop\right)^2
\end{equation}
where $\vect{S}^{(s)}$ are the spin $s$ angular momentum matrices. Eq.~(\ref{spin1Hint}) is the sum of a density-density and a spin-spin interaction. For spin 2 we have~\cite{ciobanu2000,koashi2000}
\begin{equation}\label{spin2Hint}
\cH_{\mathrm{int}}=\frac{c_0}{2}\left(\phdop\phop\right)^2+\frac{c_1}{2}\left(\phdop\vect{S}^{(2)}\phop\right)^2+\frac{c_2}{10}\left|\phop\cdot\phop\right|^2
\end{equation}
Here $\phop\cdot\phop=\sum_{m=-s}^s\left(-1\right)^{s+m}\phop_m\phop_{-m}$ is a scalar representing the amplitude of singlet pairs of spin s (Eq.~(\ref{spin1Hint}) can be expressed using this operation instead of the spin-spin interaction. In the spin 2 case both terms are needed). Our use of $c_2$ for two different quantities matches the notation of the works cited above, where one may find explicit expressions for $c_i$ $i=0,1,2$ in terms the constants $g_S$ defined above. 

In the spin 1 case it is fairly clear how to minimize Eq.~(\ref{spin1Hint}) at fixed density. For the spin 2 case things are less obvious. In the next section we will discuss a method of parametrizing the spinor $\phi$ to make this operation transparent.


\section{Identifying the spin degrees of freedom}\label{sec:spin}

In this section we will introduce the Majorana (or \emph{stellar}) representation of spin states. This provides a vivid way to picture spin ordering in higher spin condensates in which rotational symmetry is manifest. None of the calculations of Section~\ref{sec:low} depend upon this representation; its use is rather in providing a concrete way to picture the ground state manifold.

Before beginning it is worth setting the scene with a more pedestrian discussion of the spin 1 case~\cite{ho1998,ohmi1998}.

\subsection{Phases of the spin 1 gas}\label{sec:spin1phases}

Let us minimize Eq.~(\ref{spin1Hint}) with a spinor $\phi$ normalized to unity (thus we are adopting units in which the density $\rho=1$). This is a matter of minimizing (maximizing) $\phi^\dagger\vect{S}^{(1)}\phi$ for $c_2>0$ ($c_2<0$). One way to make the resulting states more clear is to write the complex spinor $\phi=\vect{a}+i\vect{b} $ where $\vect{a}$ and $\vect{b}$ are two real vectors satisfying $\vect{a}^2+\vect{b}^2=1$, and to work in cartesian components. The relationship to the usual components $\phi_m$ $m=-1,0,1$ is
\begin{eqnarray}\label{spin1cartesian}
\phi_x&=&\frac{1}{\sqrt{2}}\left(\phi_1-\phi_{-1}\right)\nonumber\\
\phi_y&=&\frac{i}{\sqrt{2}}\left(\phi_1+\phi_{-1}\right)\nonumber\\
\phi_z&=&\phi_0
\end{eqnarray}
In this basis the angular momentum matrices take the form $\left(S_i^{(1)}\right)_{jk}=-i\epsilon_{ijk}$.Then we have
\[\phi^\dagger\vect{S}^{(1)}\phi=2\vect{a}\times\vect{b}\]
For $c_2<0$ the interaction energy is minimized for $\vect{a}$ and $\vect{b}$ perpendicular and equal in magnitude. This state is termed \emph{ferromagnetic} as it corresponds to maximal polarization of the spin. The resulting order parameter manifold corresponds to the set of all configuration of a pair of orthogonal vectors, and is thus identified with the group of rotations $SO(3)$

For $c_2>0$ $\vect{a}$ and $\vect{b}$ are aligned. The resulting \emph{polar} state (the name originates from an analogous state in superfluid $^{3}$He) can therefore be written as
\begin{equation}\label{polar_parm}
\phi_{\vect{n},\theta}=e^{i\theta}\vect{n}
\end{equation}
for $\vect{n}$ a \emph{real} unit vector. Note that this parametrization has some redundancy in that the points $\left(\theta+\pi,-\vect{n}\right)$ and $\left(\theta,\vect{n}\right)$ are identified. The resulting manifold is known as the \emph{mapping torus} of the antipodal map of the sphere $S^2$.

The global topology of the above ferromagnetic and polar order parameter manifolds naturally determines the character of the topological defects in the ordered phases, and certain features of the ordering transitions, some of which have already been discussed in the literature~\cite{Mukerjee2006}. This is not the focus of the present work and topology will not be further discussed, even though the defect physics of the higher spin condensates promises to be highly non-trivial~\cite{makela2003,semenoff2007}.

The polar state has $\phi^\dagger_{\vect{n},\theta}\vect{S}^{(1)}\phi^{\vphantom{\dagger}}_{\vect{n},\theta}=0$. Nevertheless the above discussion makes it clear that polar ordering involves a choice of axis: the spinor $\phi_{\vect{n},\theta}$ is the $m=0$ state with respect to the axis $\vect{n}$. It is natural to ask for an operator that acquires a non-zero expectation value in the polar state. The obvious candidate is the spin 2 quadrupole tensor (or \emph{nematicity})
\begin{equation}\label{nematicity}
\cN^{(s)}_{ab}=\frac{1}{2}\left(S^{(s)}_aS^{(s)}_b+S^{(s)}_bS^{(s)}_a\right)-\frac{s(s+1)}{3}\delta_{ab}
\end{equation}
with 
\[\phi^\dagger_{\vect{n},\theta}\cN^{(1)}_{ab}\phi^{\vphantom{\dagger}}_{\vect{n},\theta}=\frac{1}{3}\delta_{ab}-n_an_b.\]
Such expressions are familiar in the study of nematic liquid crystals, where the vector $\vect{n}$ is known as the director. In the liquid crystal context the identification of $\vect{n}$ and $-\vect{n}$ without the phase factor in Eq.~(\ref{polar_parm}) (the order parameter manifold is then the real projective plane $\mathbb{RP}^2$) makes for very different defect physics, however~\cite{Chaikin2000}. Nematic ordering in solid state magnetic systems has been the subject of much experimental and theoretical work in recent years, with a good deal of uncertainty still remaining. The observation of the polar state in the spin 1 Bose gas would therefore be an important milestone.

Searching for higher spin order parameters as the spin of the gas particles increases becomes arduous. We now turn to a more convenient representation of the spin order.

\subsection{Majorana (stellar) representation}


The representation of a general spin $s$ state that (sometimes) bears his name was discovered by Majorana in 1932~\cite{Majorana1932}, and independently several times since~\cite{penrose1960,bacry1974}, though it has antecedents in 19$^\mathrm{th}$ century mathematics~\cite{Klein2003}. A very nice discussion can be found in Ref.~\onlinecite{Bengtsson2006}.

\begin{figure}
\centering\includegraphics[width=0.4\textwidth]{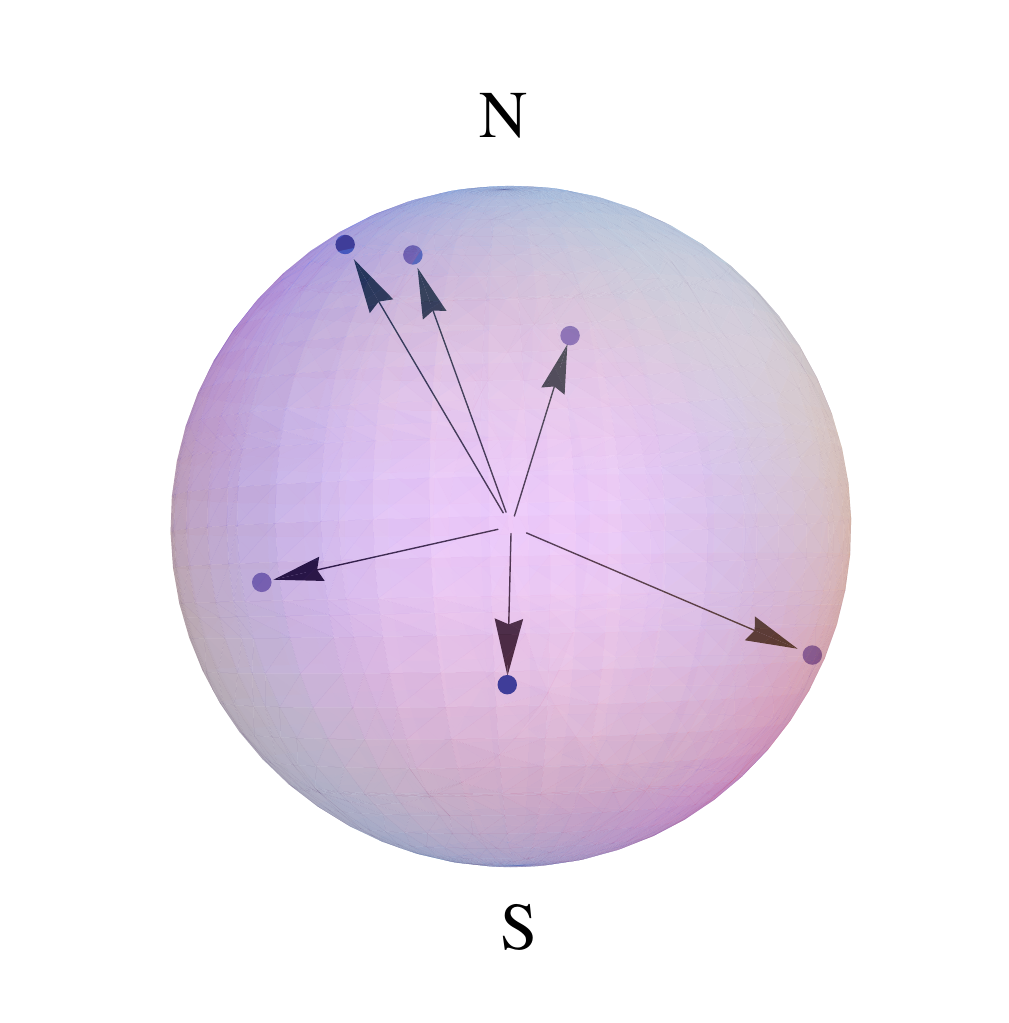}\\
\caption{(color online) Majorana representation of a $s=3$ spin state, with arrows representing the principal spinors.
\label{fig:principal}}
\end{figure}

The result is very simple to state, and represents a generalization of the Bloch sphere for spin 1/2 to arbitrary spin. Up to normalization and a phase -- thus in more mathematical terms we are parametrizing the complex projective space $\mathbb{CP}^{2s}$ -- an arbitrary spin $s$ state can be specified by locating $2s$ indistinguishable points on the unit sphere (see Fig.~\ref{fig:principal}). Such a configuration is sometimes called a \emph{constellation}, for reasons that will become clear.

There are two steps to understand why this is so. First, imagine forming our spin $s$ from $2s$ spin 1/2 in the totally symmetric subspace. An arbitrary state may then be written as a totally symmetric spinor $\Phi_{AB\cdots C}=\Phi_{(AB\cdots C)}$, where the round brackets denote the operation of symmetrization and each of the $2s+1$ indices can take the value $\uparrow$ or $\downarrow$. The relationship between $\Phi_{AB\cdots C}$ and the corresponding $2s+1$ component spinor $\phi$ is
\[\phi_m=\left(\begin{array}{c}2s \\s-m\end{array}\right)^{1/2}\Phi_{\underbrace{\uparrow\uparrow\cdots\uparrow}_{s+m}\underbrace{\downarrow\downarrow\cdots\downarrow}_{s-m}},\qquad m=-s,\ldots s.\]
Next contract every index of $\Phi_{AB\cdots C}$ with $\zeta^A=\left(\begin{array}{c}1 \\z\end{array}\right)$. Denoting by $\Phi(z)$ the resulting polynomial of order $2s$, the fundamental theorem of algebra tells us
\begin{equation}\label{majorana_poly}
\Phi(z)\equiv\Phi_{AB\cdots C}\zeta^A\zeta^B\cdots\zeta^C=\cN\prod_{i=1}^{2s}\left(z+z_i\right)
\end{equation}
with $\cN$ some normalization. Thus $\Phi_{AB\cdots C}$ may be written
\begin{equation}\label{root_spinors}
\Phi_{AB\cdots C}=\alpha_{(A}\beta_{\vphantom{(}B}\cdots \gamma_{C)}
\end{equation}
with the \emph{principal spinors} $\alpha_A$, $\beta_B$, etc. related to the $\{z_i\}$ by
\begin{eqnarray}
\frac{\alpha_\uparrow}{\alpha_\downarrow}&=&z_1\nonumber\\
\frac{\beta_\uparrow}{\beta_\downarrow}&=&z_2\nonumber\\
&\cdots&\nonumber\\
\frac{\gamma_\uparrow}{\gamma_\downarrow}&=&z_{2s}
\end{eqnarray}
relations that are unchanged if we normalize the spinors, in which case they correspond to $2s$ points on the Bloch sphere with coordinates $\{\theta_i,\varphi_i\}$, $i=1,\ldots 2s$
\begin{eqnarray*}
 (\alpha_\uparrow,\alpha_\downarrow)&=&\left(e^{i\varphi_1/2}\cos\frac{\theta_1}{2} ,e^{-i\varphi_1/2}\sin\frac{\theta_1}{2}\right)\\ 
 (\beta_\uparrow,\beta_\downarrow)&=&\left(e^{i\varphi_2/2}\cos\frac{\theta_2}{2} ,e^{-i\varphi_2/2}\sin\frac{\theta_2}{2}\right)\\
&\cdots&\\
(\gamma_\uparrow,\gamma_\downarrow)&=&\left(e^{i\varphi_{2s}/2}\cos\frac{\theta_{2s}}{2} ,e^{-i\varphi_{2s}/2}\sin\frac{\theta_{2s}}{2}\right)
\end{eqnarray*}
(notice that $\Phi_{AB\cdots C}$ in Eq.~(\ref{root_spinors}) is not in general normalized when the principal spinors are). Then (minus) the roots of $\Phi(z)$ can be written $z_i=e^{i\varphi_i}\cot\theta_i/2$ and correspond to stereographic projection from the north pole to the plane tangent to the sphere at the south pole.

The beautiful feature of the Majorana representation is that rotations act simply as rotations on the Bloch sphere. Of course, it is useful to have an explicit expression for the polynomial $\Phi(z)$ in terms of the $2s+1$ components of the spin $s$ state $\phi$. It is easy to show 
\begin{equation}\label{poly_exp}
\Phi(z)=\sum_{m=-s}^s \phi_mz^{s-m}\left(\begin{array}{c}2s \\s-m\end{array}\right)^{1/2}.
\end{equation}

If spinor indices are raised and lowered using the antisymmetric tensors $\epsilon_{AB}$ and $\epsilon^{AB}$ (with $\epsilon_{\uparrow\downarrow}=1$ and $\epsilon^{AB}=-\epsilon_{AB}$)
\[\alpha^A=\epsilon^{AB}\alpha_B\qquad \leftrightarrow\qquad \alpha_A=\alpha^B\epsilon_{BA},\]
then rotational invariance upon contraction of indices is guaranteed (in fact the result is invariant under the larger group $SL(2,\mathbb{C})$, a result that will be useful later) . If we denote by $\bar{\phi}$  the result on $\phi$ of raising all indices of the corresponding symmetric spinor, then one can readily see that $\bar\phi_m=(-1)^{s+m}\phi_{-m}$, and thus
\[\phi\cdot\psi=\sum_{m=-s}^s\phi_m\bar\psi_m=\Phi_{AB\cdots C}\Psi^{AB\cdots C}\]. 

After raising indices of the principal spinors we have $\alpha^\uparrow/\alpha^\downarrow=-1/z_1$, etc.. Under stereographic projection $z\to -1/z^*$ represents the antipodal map on the unit sphere. Thus we see that the spinor $\bar{\phi}^*$ is represented by a set of points antipodal to those representing $\phi$. Furthermore, the Majorana representation of a normalized state with $|\phi\cdot\phi|=1$, corresponding to $\phi=e^{2i\theta}\bar{\phi}^*$, consists of pairs of antipodal points (and is thus only possible for integer spin). This fact will be useful in minimizing the interaction energy (recall the form of Eq.~(\ref{spin2Hint})). The transformation $\cT:\phi_m\to (\cT\phi)_m=\bar\phi^*_m$ is in fact the (anti-unitary) operation of time reversal.

\subsection{Spin ordering in the Majorana representation}

The use of the Majorana representation to visualize spin ordering in a Bose gas was suggested in Ref.~\onlinecite{barnett2006a}. Let us first see how the phases of the spin 1 gas discussed in Section~\ref{sec:spin1phases} appear in this representation, before moving on to the spin 2 case.

\subsubsection{Spin 1}

As mentioned in Section~\ref{sec:Hint}, Eq.~(\ref{spin1Hint}), the interaction Hamiltonian in the spin 1 case may be written
\begin{equation}\label{spin1Hint_rewrite}
\cH_{\mathrm{int}}=\frac{c_0+c_2}{2}\left(\phdop\phop\right)^2-\frac{c_2}{2} \left|\phop\cdot\phop\right|^2
\end{equation}
For $c_2>0$ we should maximize $\left|\phop\cdot\phop\right|^2$. Based on the discussion of the previous section, this corresponds to placing the two points antipodally in the Majorana representation. It is evident that the corresponding spin 1 spinor is just the symmetric $m=0$ state with respect to the resulting axis. This is just the polar state described before.

For $c_2<0$ $\left|\phop\cdot\phop\right|^2$ can be set to zero by making the two principal spinors equal since $\alpha_A\alpha^A=0$. It is clear that this represents the ferromagnet, being a maximally polarized (or \emph{coherent}) spin state, a result that generalizes to arbitrary $s$.

\subsubsection{Spin 2} \label{sec:spin2phases}

\begin{figure}
\centering\includegraphics[width=0.3\textwidth]{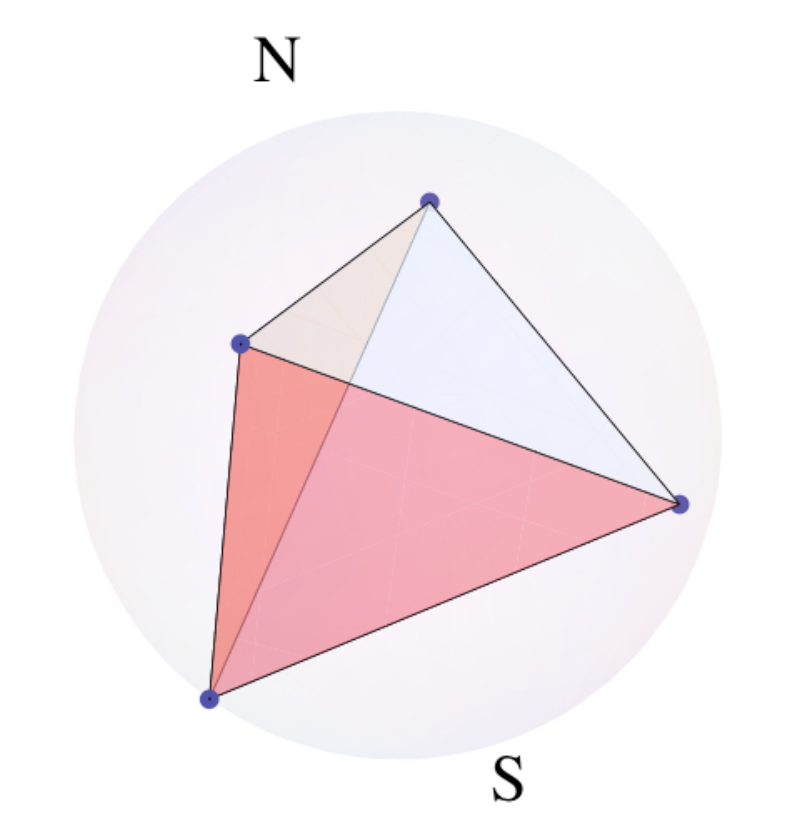}\\
\includegraphics[width=0.2\textwidth]{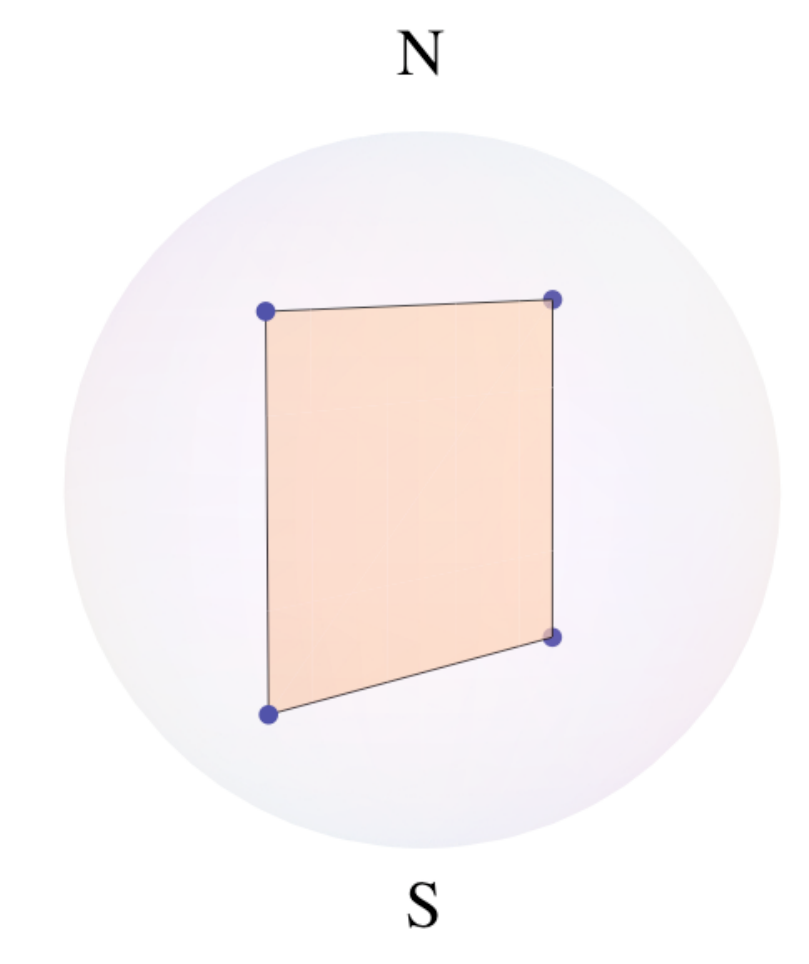}
\includegraphics[width=0.2\textwidth]{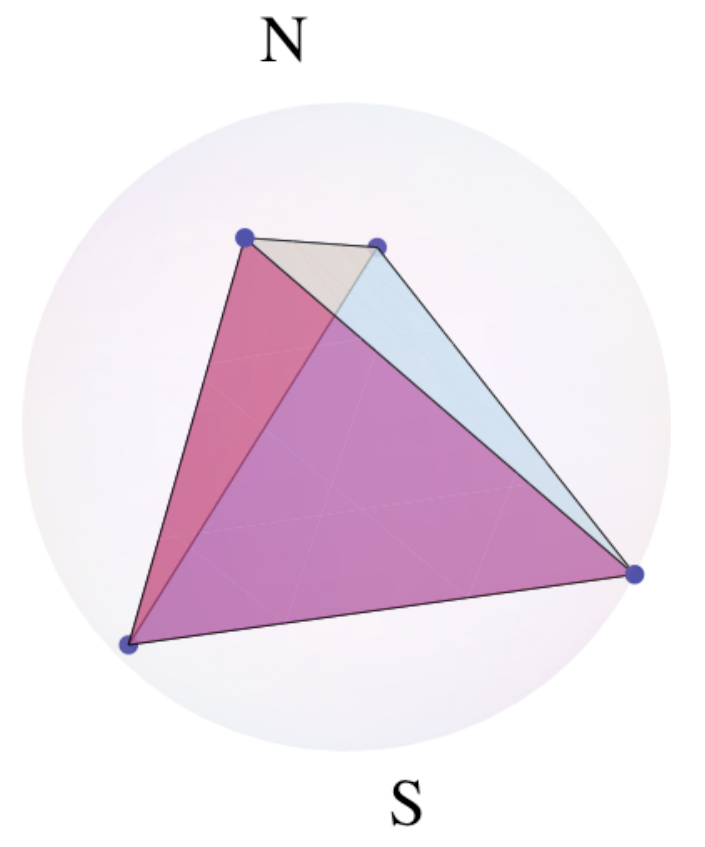}
\caption{(color online) Top: disphenoid corresponding to the Majorana representation of the state with $\chi=\pi/4$, $\eta=\pi/2$ in Eq.~(\ref{disphenoid}). Bottom left: rectangular state with $\chi=0$, $\eta=\pi/2$. Bottom right: tetrahedron with $\chi=\eta=\pi/2$.
\label{fig:disphenoid}}
\end{figure}

Suppose that in Eq.~(\ref{spin2Hint}) $c_1>0$. The corresponding term in the Hamiltonian can be fully satisfied by states with $\phi^\dagger \vect{S}^{(2)}\phi=0$. By inspecting the spin 2 angular momentum matrices
\begin{widetext}
\begin{equation}\label{spin2}
S^{(2)}_x=\left(\begin{array}{ccccc}0 & 1 & 0 & 0 & 0 \\1 & 0 & \sqrt{3/2} & 0 & 0 \\0 & \sqrt{3/2} & 0 & \sqrt{3/2} & 0 \\0 & 0 & \sqrt{3/2} & 0 & 1 \\0 & 0 & 0 & 1 & 0\end{array}\right),\,
S^{(2)}_y=i\left(\begin{array}{ccccc}0 & -1 & 0 & 0 & 0 \\1 & 0 & -\sqrt{3/2} & 0 & 0 \\0 & \sqrt{3/2} & 0 & -\sqrt{3/2} & 0 \\0 & 0 & \sqrt{3/2} & 0 & -1 \\0 & 0 & 0 & 1 & 0\end{array}\right),\,
S^{(2)}_z=\left(\begin{array}{ccccc}2 & 0 & 0 & 0 & 0 \\0 & 1 &0& 0 & 0 \\0 &0 & 0 & 0 & 0 \\0 & 0 & 0 & -1 & 1 \\0 & 0 & 0 & 0 & -2\end{array}\right),
\end{equation}
\end{widetext}
it is easy to convince oneself that up to rotation the most general state satisfying these conditions is 
\begin{equation}\label{disphenoid}
\phi_{\eta,\chi}=\left(\begin{array}{c}\frac{e^{-i\chi/2}\sin\eta/2}{\sqrt{2}} \\0 \\e^{i\chi/2}\cos\eta/2 \\0 \\\frac{e^{-i\chi/2}\sin\eta/2}{\sqrt{2}}\end{array}\right).
\end{equation}
The principal spinors form a polyhedron with four identical triangular faces known as a \emph{disphenoid} (see Fig.~\ref{fig:disphenoid}).
To fix the parameters $\eta$, $\chi$ in Eq.~(\ref{disphenoid}) we turn to the third term in Eq.~(\ref{spin2Hint}). If $c_2<0$, this term is minimized by placing four points in two antipodal pairs so that $|\phi\cdot\phi|^2=1$. The principal spinors form a rectangle, corresponding to $\chi=0$ in Eq.~(\ref{disphenoid}). The aspect ratio of the rectangle varies with $\eta$, with $\eta=\left(2n+1\right)\pi/3$ being a square and $\eta=0$ corresponding to a pair of points at either pole. The fact that the energy is minimized for \emph{any} $\eta$ in this parameter regime is rather surprising, and we will return to it briefly below. This state will be referred to as \emph{rectangular} in the following (in Ref.~\onlinecite{barnett2006a} it was called nematic since $\cN_{ab}\neq 0)$.

For $c_2>0$ $|\phi\cdot\phi|^2$ should be minimized. This can be done by taking $\chi=\eta=\pi/2$. The result is a regular tetrahedron. Note that this state has $\cN_{ab}=0$. 

For $c_1<0$ one must compare the energy of the rectangular state with that of the ferromagnet. The resulting phase diagram can be found in Refs.~\onlinecite{ciobanu2000,barnett2006a}. Note that while the rectangular state maximizes the magnitude of the quadratic scalar
\begin{equation}\label{quad_inv}
I\equiv\phi\cdot\phi=\Phi^{ABCD}\Phi_{ABCD},
\end{equation}
the tetrahedral state maximizes the cubic invariant 
\begin{equation}\label{cubic_inv}
J\equiv \Phi^{AB}_{CD}\Phi^{CD}_{EF}\Phi^{EF}_{AB}.
\end{equation}
The inclusion of a sextic term proportional to $|J|^2$ in the interaction energy would lift the accidental degeneracy in the $\eta$ parameter discussed above. A microscopic derivation of such a term is discussed in Refs.~\onlinecite{turner2007,song2007a}, with a positive sign favoring the square state ($\eta=\left(2n+1\right)\pi/3$) and a negative sign the uniaxial state in which two pairs of points coincide ($\eta=0$). The inclusion of such a term in our formalism is a straightforward matter, and we will not discuss it further.


The rectangular and tetrahedral states evidently have certain discrete symmetries that are rather hard to discern by inspection of Eq.~(\ref{disphenoid}), and indeed went unnoticed in the earliest works on the spin 2 condensate~\cite{ciobanu2000,ueda2002}. It appears that the term `cyclic' used in several works to describe the tetrahedral phase is  a consequence of a misidentification of the symmetry. This illustrates the utility of the Majorana representation in the visualization of spin order. 

 
\subsection{Ground state manifolds}\label{gs_man}

\subsubsection{Global structure}\label{sec:global}

In Section~\ref{sec:spin1phases} we identified the order parameter manifolds of the phases of the spin 1 gas. With the help of the Majorana representation we can now do the same for the spin 2 case. Roughly speaking, we expect the manifold to consist of all configurations related to those of Section~\ref{sec:spin2phases} by rotation. Some rotations will leave the configuration of principal spinors unchanged, however, so the manifold cannot simply be identified with the rotation group $SO(3)$. In mathematical terms the problem is to determine the \emph{orbits} of a reference spinor under the action of the spin $s$ representation of the rotation group. 

If we ignore the phase of the spinor for a moment, so that we are considering orbits in $\mathbb{CP}^{2s}$, this problem can be solved using the Majorana representation by considering the manifold of constellations generated by all possible rotations~\cite{bacry1974}.
For instance, the ferromagnetic spin 1 state has orbit in $\mathbb{CP}^2$ equal to $S^2$ (and this is true in general for any spin $s$ coherent state), while for the polar state we have $S^2$ with antipodal points identified: the real projective plane $\mathbb{RP}^2$. In general if one finds a configuration of principal spinors unchanged under some subgroup of $\Gamma\subset SO(3)$ (the \emph{stabilizer} subgroup of a constellation), then the orbit is given by $SO(3)/\Gamma$. Thus in the case of spin 1, $\Gamma=SO(2)$ for the ferromagnet and $\Gamma=O(2)$ in the polar phase, since in the latter case a parity transformation also leaves the points unchanged.

The reader will notice that these are \emph{not} the order parameter manifolds identified in Section~\ref{sec:spin1phases} for the spin 1 case. We have neglected the phase of the spinor, which is a real degree of freedom. We might then guess that any spinor on the order parameter manifold can be written
\begin{equation}\label{gen_param}
\phi_{R,\theta}=e^{i\theta}D^{(s)}(R)\phi_0
\end{equation}
where $D^{(s)}(R)$ is the spin $s$ representation of the rotation $R$ and $\phi_0$ is some reference spinor corresponding to the phase in question. This does not mean that the order parameter manifold is $SO(3)\times U(1)$, because the stabilizer subgroups mentioned above leave the spinor corresponding to a particular constellation unchanged up to a phase. We denote these phases as $\Lambda(\gamma)$, with $\gamma\in \Gamma$ . They must form a one-dimensional unitary representation of $\Gamma$
\[\Lambda(\gamma_1\gamma_2)=\Lambda(\gamma_1)+\Lambda(\gamma_2)\qquad \mathrm{mod}\,2\pi.\]
and allow us to make the identification 
\begin{equation}\label{man_ident}
\phi_{R,\theta}=\phi_{R \gamma ,\theta-\Lambda(\gamma)}\qquad \gamma\in\Gamma,
\end{equation}
showing that the order parameter manifold is $\frac{SO(3)\times U(1)}{\tilde\Gamma}$, where the tilde is to denote the action of $\Gamma$ on $SO(3)\times U(1)$: $(R,\theta)\to ( R\gamma,\theta-\Lambda(\gamma))$ 

The simplest case to consider is the polar phase with $\Gamma=O(2)$, for which the only non-trivial phase is a $-1$ associated with the parity transformation. If $\Gamma$ were $SO(2)$ with only trivial phases, the ground state manifold would be $S^1\times S^2$, with the first factor coming  from the $U(1)$ and the second from $SO(3)/SO(2)$. The parity transformation and the associated minus sign are responsible for the identification $\left(\theta+\pi,-\vect{n}\right)=\left(\theta,\vect{n}\right)$  already discussed in Section~\ref{sec:spin1phases}.

\begin{figure}
\includegraphics[width=0.4\textwidth]{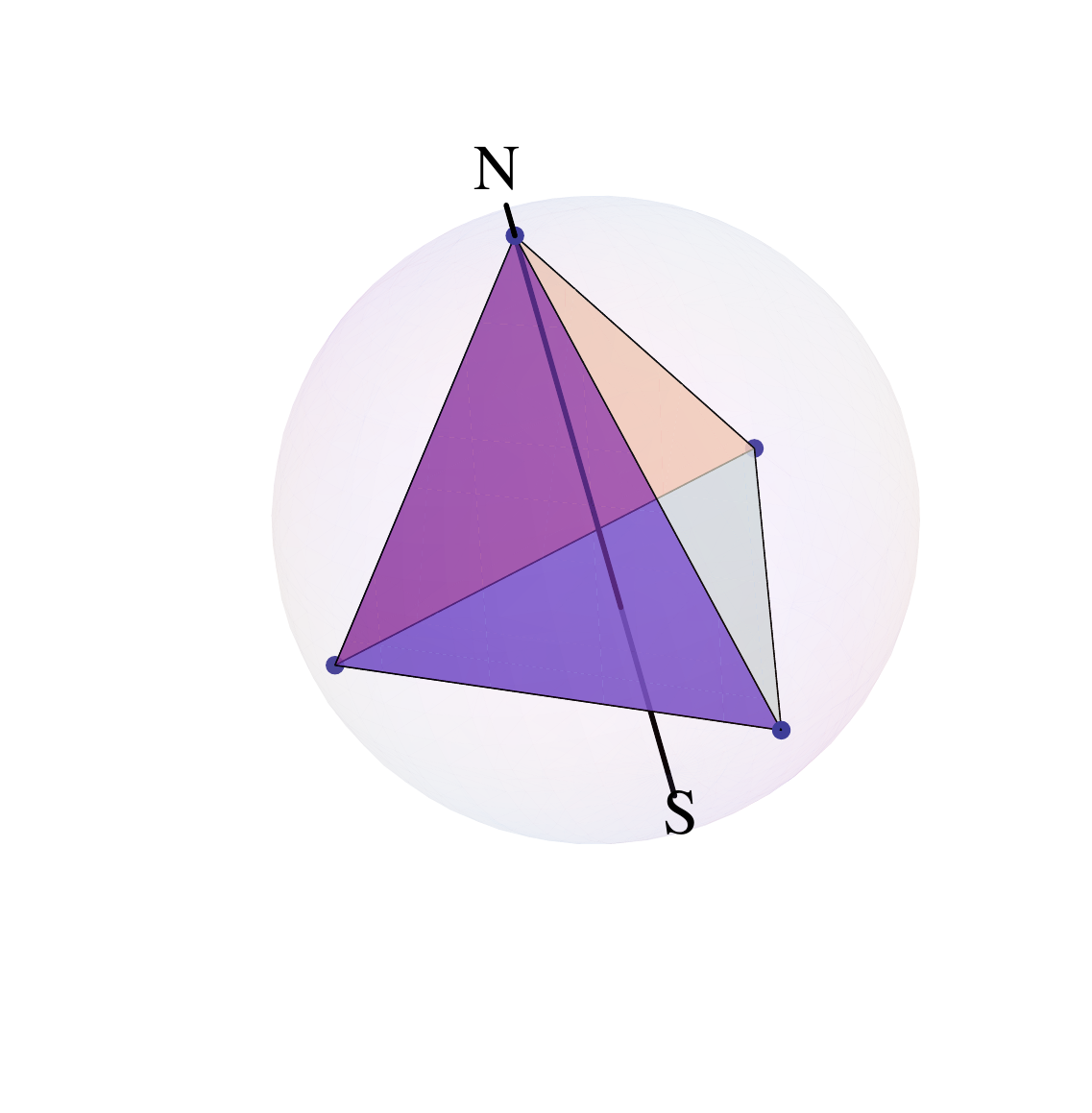}
\caption{(color online) Tetrahedron with 3-fold axis marked, see Eq.~(\ref{3axis}).
\label{fig:3axis}}
\end{figure}

To turn to a less trivial example, let us see how this works for the case of the tetrahedral phase. In this case $\Gamma=T$, the symmetry group of the tetrahedron. A tetrahedron has 3 orthogonal 2-fold axes, and 4 3-fold axes. For the representative spinor given earlier
\begin{equation}\label{2axis}
\phi=\left(\begin{array}{c}\frac{\sqrt{-i}}{2} \\0 \\\sqrt{\frac{i}{2}} \\0 \\\frac{\sqrt{-i}}{2}\end{array}\right),
\end{equation}
the $z$ axis is aligned with one of the 2-fold axes. One easily verifies that $z$ axis rotations through $\pi$ leave the spinor unchanged.   Alternatively, we can align one of the 3-fold axes with the $z$ axis with the choice (see Fig.~\ref{fig:3axis})
\begin{equation}\label{3axis}
\phi=\left(\begin{array}{c}0 \\\sqrt{\frac{2}{3}} \\0 \\0 \\\sqrt{\frac{1}{3}}\end{array}\right),
\end{equation}
(this is most easily seen by considering the Majorana polynomial, which has a root at $z=0$). Now a rotation through $\pm2\pi/3$ is seen to reproduce the same spinor but with phase factors $e^{\pm2\pi i/3}$. It is not hard to verify that these phases form a one-dimensional representation of $T$ (the other non-trivial one-dimensional representation comes from changing the sense of the 3-fold axes). The topological properties of the resulting space $\frac{SO(3)\times U(1)}{\tilde T}$, and the implications for superfluid vortices in the tetrahedral phase, were discussed in Ref.~\onlinecite{semenoff2007}.

For the rectangular phase $\Gamma=D_2$ in general, but $D_4$ for the square case ($\eta=\left(2n+1\right)\pi/3$). Here $D_n$ denotes the dihedral group. The only non-trivial phases occur in the latter case, as may be seen by considering the value $\eta=\pi$, when the four points lie on the equator of the Bloch sphere. Then we have
\begin{equation}\label{Square4axis}
\phi=\left(\begin{array}{c}\frac{1}{\sqrt{2}} \\0 \\0 \\0 \\\frac{1}{\sqrt{2}}\end{array}\right),
\end{equation}
and a $\pm\pi/2$ rotation about the $z$ axis is seen to give rise to a $-1$. 

This construction generalizes readily to other ordered states of arbitrary spin once the corresponding stabilizer subgroups and phases are identified~\cite{barnett2006a}. Note that when $\Gamma$ is discrete, as for the spin 2 phases other than the ferromagnet, the order parameter manifold is 4 dimensional. 




\subsubsection{Local geometry}

Next we turn to the local properties of the order parameter manifold. The inner product naturally endows this space with a metric~\cite{provost1980}. Using the parametrization Eq.~(\ref{gen_param}) consider two states $\phi_{R,\theta}$ and $\phi_{R',\theta'}$ related $\theta'=\theta+d\theta$ and $R'=RR^\psi$ with $R^{\psi}_{ab}=\delta_{ab}-\psi_c\epsilon_{abc}$ an infinitesimal rotation corresponding to
\[D^{(s)}(R^\psi)=\openone-i\bm{\psi}\cdot\vect{S}^{(s)}.\]
We find the squared distance between these two states to be
%
\begin{equation}\label{distance}
||\phi_{R',\theta'} -\phi_{R,\theta}||^2=
d\theta^2+\psi_a\psi_b g_{ab}-2d\theta\bm{\psi}\cdot \left(\phdop_0 \vect{S}^{(s)}\phop_0\right)
\end{equation}
%
where the tensor $g_{ab}$ is 
\begin{equation}\label{metric}
g_{ab}=\frac{1}{2}\phdop_0\{S_a^{(s)},S_b^{(s)}\}\phop_0.
\end{equation}
Eq.~(\ref{distance}) makes it clear that if $\phdop_0\vect{S}^{(s)}\phop_0\neq 0$, rotations and phase changes are coupled together. There is an arbitrariness in the way the phase is apportioned between the two factors in Eq.~(\ref{gen_param}) that amounts to a choice of gauge. This gauge structure was discussed in Ref.~\onlinecite{Lamacraft2008} for the case of the ferromagnet, the only one of the phases discussed in Section~\ref{sec:spin} with $\phdop_0\vect{S}^{(s)}\phop_0\neq 0$.  The discussion of the general case is relegated to Appendix~\ref{sec:gauge}. Our main interest is in the other phases having $\phdop_0\vect{S}^{(s)}\phop_0= 0$, for which Eq.~(\ref{distance}) decouples into separate contributions from the change in phase and the rotation, with the latter being characterized by the metric tensor $g_{ab}$. Note that $g_{ab}$ is simply related to the nematicity $\cN_{ab}$ in Eq.~(\ref{nematicity}). The notion of distance described by the metric tensor is left invariant i.e. preserved if states $\phi_{R,\theta}$ are mapped by $R\to \tilde R R$ for some $\tilde R$, but not right invariant, under which $g\to \tilde R g \tilde R^T$.


Focusing now on the spin 2 case, we evaluate the metric for the state Eq.~(\ref{disphenoid})
\begin{widetext}
\begin{equation}\label{disphenoid_metric}
g=\left(\begin{array}{ccc}2+\cos\eta+\sqrt{3}\cos\chi\sin\eta & 0 & 0 \\0 & 2+\cos\eta-\sqrt{3}\cos\chi\sin\eta & 0 \\0 & 0 & 4\sin^2\frac{\eta}{2}\end{array}\right)
\end{equation}
\end{widetext}
%
For the tetrahedral phase ($\chi=\eta=\pi/2$) $g=2\times\openone$, showing that the order parameter manifold has a left and right invariant geometry. This is perhaps not surprising given the highly symmetric arrangement of points in the Majorana representation. For the rectangular phase ($\chi=0$), we see that in the case $\eta=0$, $g=\mathrm{diag}\left(3,3,0\right)$. The physical meaning is clear: because we have two points at either pole the stabilizer subgroup is $O(2)$, so that the order parameter manifold is only three dimensional. The same holds true for the spin 1 polar phase.

We will see that the metric plays a crucial role in fixing the dynamics on the order parameter manifold.





\subsection{Conjugate variables}

Having characterized the order parameter manifold for the spinor condensates, we are almost ready to study the dynamics on that manifold. It remains to identify the \emph{conjugate variables}. We expect these to be coupled by the first term of Eq.~(\ref{GP_action}), which expresses the conjugacy of $\phi$ and $\phi^{\dagger}$. Substituting the parametrization Eq.~(\ref{gen_param}) into that term gives
\begin{eqnarray}\label{conj_param}
i\phdop\partial_t\phop&=&-\partial_t\theta+i\phdop_0\left(D^{(s)\dagger}\partial_t D^{(s)}\right)\phop_0\nonumber\\
&=&-\partial_t\theta+\phdop_{0}\bm{\omega}_t \cdot \vect{S}^{(s)}\phop_{0}\nonumber\\
\end{eqnarray}
where $\omega_{t,a}=\frac{1}{2}\epsilon_{abc}\left(R^T\partial_t R^{\vphantom{T}}\right)_{cb}$ are the components of the angular velocity. By analogy with rigid body dynamics, we refer to this as the `body frame' angular velocity: Eq.~(\ref{conj_param}) shows that it is conjugate to the `body frame' angular momentum $\phdop_0\vect{S}^{(s)}\phop_0$, i.e. that of the unrotated state. As is well known, the time integral of the term $i\phdop\partial_t\phop$ has an alternative interpretation as the Berry phase associated with the time evolution of $\phop$. In this context the formula Eq.~(\ref{conj_param}) appears in Ref.~\onlinecite{hannay1998}.

\begin{figure}
\includegraphics[width=0.23\textwidth]{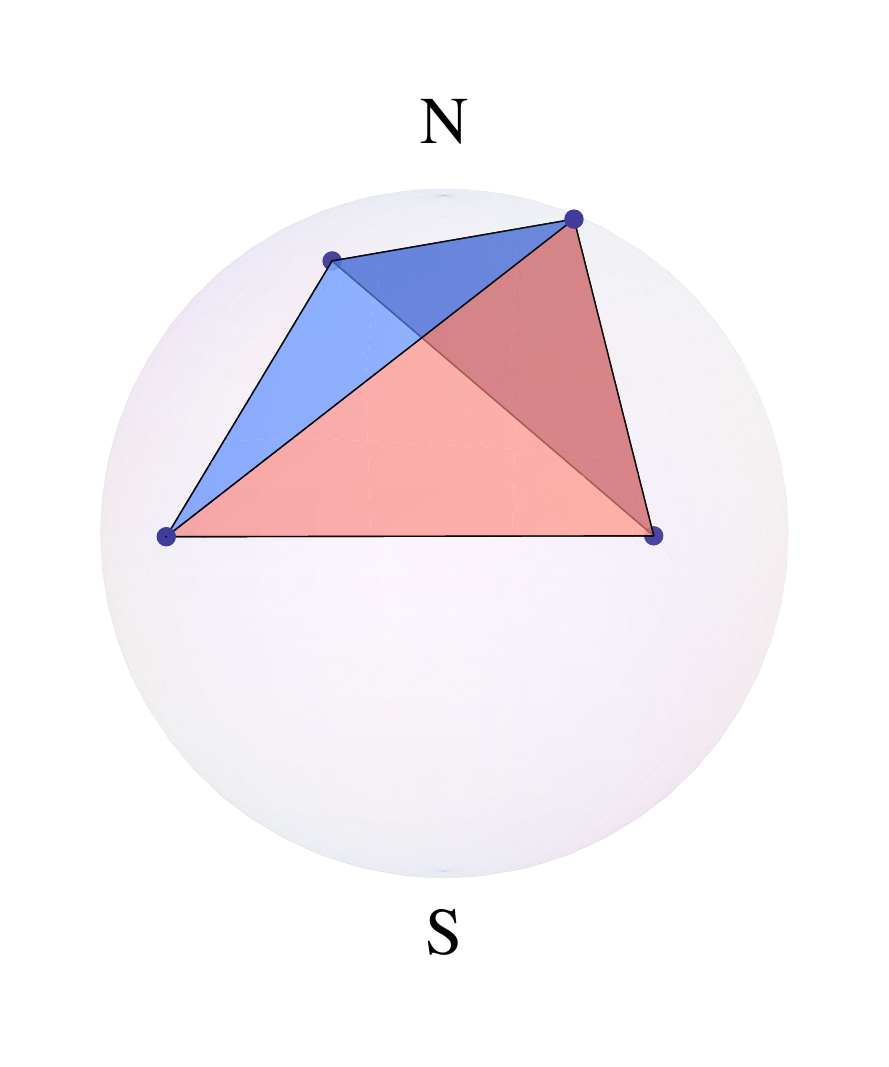}
\includegraphics[width=0.23\textwidth]{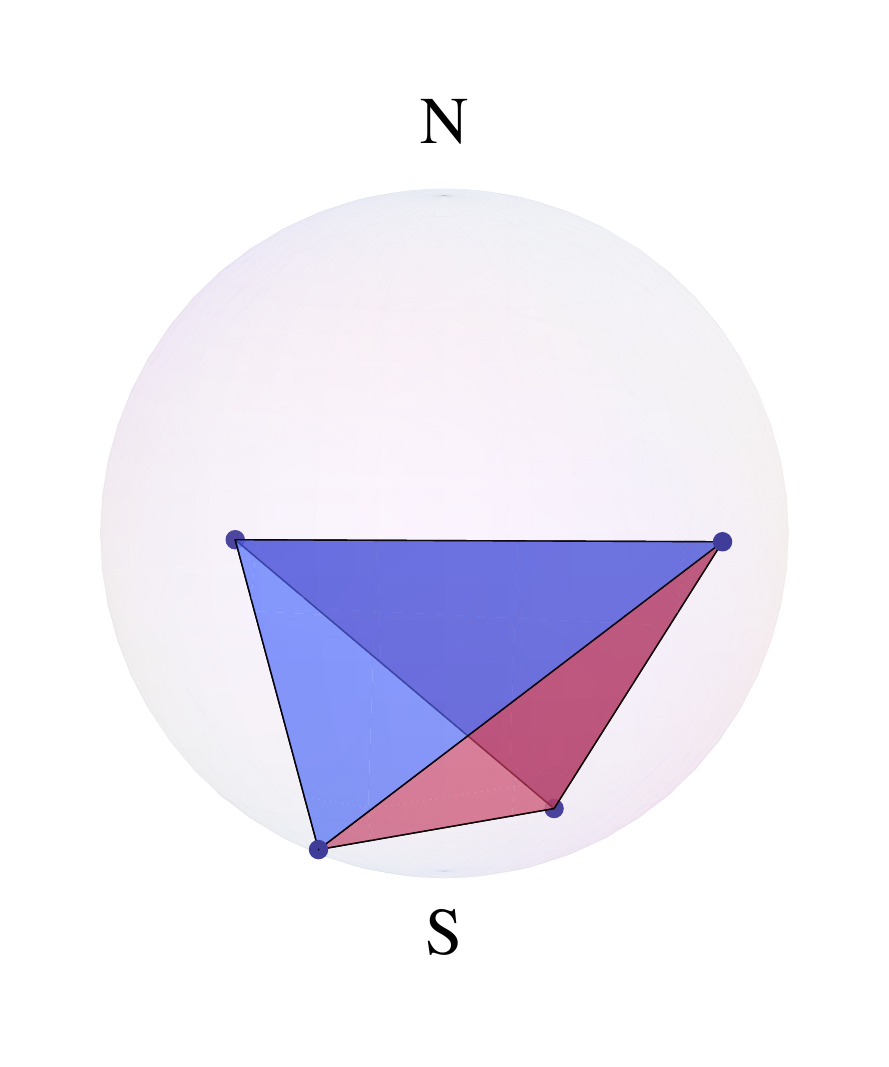}
\caption{(color online) Distortion of the tetrahedral state shown in Fig.~\ref{fig:disphenoid} due to boosts in the $z$-direction
\label{fig:distort}}
\end{figure}

Thus in the cases of interest where $\phdop_0\vect{S}^{(s)}\phop_0=0$ the variables conjugate to the rotations are nonzero only as one deviates from the order parameter manifold. To account for these deviations we generalize the parametrization Eq.~(\ref{gen_param}) to
\begin{equation}\label{gen_param_conj}
\phi_{R,\vect{l},\theta}=e^{i\theta}D^{(s)}(R)\phi_{\vect{l}}
\end{equation}
where $\phi_{\vect{l}}$ is a state with $\phdop_{\vect{l}}\vect{S}^{s}\phop_{\vect{l}}=\vect{l}$. Note that
\[\phdop_{R,\vect{l},\theta}\vect{S}^{(s)}\phop_{R,\vect{l},\theta}=R\vect{l}\]
$R\vect{l}$ is angular momentum in the `lab frame'. To find the deviation corresponding to $\phi_{\vect{l}}$, let us consider the state
\begin{equation}\label{l_deform}
\phi_{\vect{l}}=\cN(\vect{l})B^{(s)}(\vect{l})\phi_0
\end{equation}
where $\cN(\vect{l})$ is some normalization factor, and 
\[B^{(s)}(\vect{l})=\exp\left(\frac{1}{2}\left(g^{-1}\right)_{ab}l_a S^{(s)}_b\right).\]
It is not hard to see that to quadratic order the normalization takes the form
\begin{equation}\label{norm}
\cN(\vect{l})=1-\frac{1}{4}\left(g^{-1}\right)_{ab}l_al_b+\cdots
\end{equation}
and that to this order $\phdop_{\vect{l}}\vect{S}^{s}\phop_{\vect{l}}=\vect{l}$, as required. The effect of the $\vect{l}$-distortion on the tetrahedral state is shown in Fig.~\ref{fig:distort}.

Using the parametrization Eq.~(\ref{gen_param_conj}) in Eq.~(\ref{conj_param}) gives then
\begin{equation*}\label{conj_param_new}
i\phdop\partial_t\phop=-\partial_t\theta+\bm{\omega}\cdot\vect{l},
\end{equation*}
to first order in $\vect{l}$.

We make the following aside. The matrix $D^{(s)}(R)B^{(s)}(\vect{l})$ that acts on $\phi_0$ is a polar decomposition of the $D_{s,0}$ representation of an element of $SL(2,\mathbb{C})$~\cite{richtmyer1978}. Now $SL(2,\mathbb{C})/\mathbb{Z}_2$ is isomorphic to the connected Lorentz group. This isomorphism has a beautiful physical counterpart. The Lorentz transformations have a natural action on the Celestial sphere, the space of light rays on the past (say) light cone. Further, the elements of $SL(2,\mathbb{C})$
\[A=\left(\begin{array}{cc}a & b \\c & d\end{array}\right),\qquad ad-bc=1,\]
have a natural action on the Bloch sphere: $\left(\begin{array}{c}\alpha_\uparrow \\\alpha_\downarrow\end{array}\right)\to A\left(\begin{array}{c}\alpha_\uparrow \\\alpha_\downarrow\end{array}\right)$, corresponding to a M\"obius transformation on the stereographic coordinates $z=\alpha_\uparrow/\alpha_\downarrow$
\begin{equation}\label{mobius}
z\to \frac{az+b}{cz+d}.
\end{equation}
(the fact that $A$ and $-A$ correspond to the same M\"obius transformation accounts for the $\mathbb{Z}_2$ quotient above). Remarkably, group elements related by the isomorphism mentioned above correspond to identical transformations of the sphere~\cite{bacry2004}! The deformations  of  Eq.~(\ref{l_deform}) are just those generated on the Celestial sphere by a boost of the frame of reference, which in particular determines the aberration of the fixed stars. Thus the name `stellar representation' sometimes used to describe the Majorana picture of spin states is more than picturesque. 
 
With the parametrization Eq.~(\ref{gen_param_conj}) we have accounted for six spin degrees of freedom (plus the superfluid phase): three rotation variables parametrizing the order parameter manifold and three conjugate variables. The remaining $4s-6$ variables required to specify the spin state do not correspond to any broken symmetries and must describe gapped modes. 

For the spin 2 case, the missing two degrees of freedom can be readily found by considering the ratio of the two (complex) $SL(2,\mathbb{C})$  invariants  in Eqns.~(\ref{quad_inv},\ref{cubic_inv}).
\begin{equation}\label{mu_inv}
\mu=\frac{I^3}{J^2}
\end{equation}
The powers are chosen so that the normalization in Eq.~(\ref{l_deform}) drops out. By construction $\mu$ is unchanged for all $\vect{l}$ and $R$ starting from the state Eq.~(\ref{disphenoid}), and could be expressed in terms of $\eta$ and $\chi$. 

This concludes our discussion of the spin degrees of freedom in a spinor condensate. With this background, we will see that the derivation of the low energy Lagrangian is extremely straightforward.

\section{Dynamics near the order parameter manifold} \label{sec:low}

\subsection{Low energy Lagrangian}

We are going to use the parametrization Eq.~(\ref{gen_param_conj}) in the Lagrangian Eq.~(\ref{GP_action}). In doing so, we are treating the problem as as \emph{constrained} dynamical system, in which deviations from the ground state manifold associated with the gapped modes described in the previous section are assumed to be infinitely stiff. The Lagrangian for the spin degrees of freedom then takes the form
\begin{equation}\label{L_exp}
\cL_{\mathrm{spin}}=\bm{\omega}_t\cdot\vect{l}-\frac{1}{2}g_{ab}\omega_{i,a}\omega_{i,b}-\cH_{\mathrm{int}}(\vect{l})
\end{equation}
where $\omega_{i, a}=\frac{1}{2}\epsilon_{abc}\left(R^T\partial_i R^{\vphantom{T}}\right)_{cb}$  $i=x,y,z$, in analogy to the earlier definition of $\omega_{t,a}$. We could additionally allow for a variation $\delta\rho$ in the density of the gas, which is conjugate to the phase $\theta$, to describe the density modes, but will not do so here. In Eq.~(\ref{L_exp}) we have not included the gradient of the conjugate variables (including $\delta\rho$) in the part arising from the kinetic energy, as such terms can be neglected in the long wavelength limit of interest. The conjugate variables do however appear in the interaction term
\begin{equation}\label{Hint_quad}
\cH_{\mathrm{int}}(\vect{l})=\frac{1}{2}\left(I^{-1}\right)_{ab}l_al_b
\end{equation}
The notation is of course chosen to emphasize the rigid body analogy. The precise form of the `inertia tensor' will depend on the phase under consideration; we discuss the spin 2 phases for definiteness. In that case the interaction Hamiltonian has the form Eq.~(\ref{spin2Hint}). Computing the quadratic variation of this expression with the conjugate variables is facilitated by the $SL(2,\mathbb{C})$ invariance of the third term: its variation is determined solely by the normalization in Eq.~(\ref{norm}). We obtain
\begin{eqnarray}\label{spin2_inertia}
\left(I^{-1}\right)_{ab}=
\begin{cases}
c_1\delta_{ab} & \text{tetrahedral phase} \\
c_1\delta_{ab}-\frac{c_2}{5}\left(g^{-1}\right)_{ab} & \text{rectangular phase} 
\end{cases}
\end{eqnarray}
It is then straightforward to eliminate the $\vect{l}$ degrees of freedom using the equation of motion, obtained from Eq.~(\ref{L_exp})
\[l_a=I_{ab}\omega_{t,b}\]
to obtain the final result
\begin{equation}\label{L_exp_l_elim}
\cL_{\mathrm{spin}}=\frac{1}{2}I_{ab}\omega_{t,a}\omega_{t,b}-\frac{1}{2}g_{ab}\omega_{i,a}\omega_{i,b}
\end{equation}
Eq.~(\ref{L_exp_l_elim}) represents the main conclusion of this work, being the low energy Lagrangian for the spin degrees of freedom of the condensate. It takes the form of a  sigma model in $3+1$ dimensions with target space $SO(3)$ (if we ignore the possibility of vortices in the superfluid phase $\theta$ the subtle global structure of the ground state manifold discussed in Section~\ref{sec:global} can be ignored). In the case of the tetrahedral phase, the metric tensor $g=2\openone$, and the Lagrangian Eq.~(\ref{L_exp_l_elim}) becomes that of the \emph{principal chiral model}, having independent left and right $SO(3)$ symmetries.

An alternative form for Eq.~(\ref{L_exp_l_elim}) follows from noting that, if $M=\mathrm{diag}\left(m_1,m_2,m_3\right)$
\[M_{ab}\omega_{\mu,a}\omega_{\mu,b}=\tr\left[\tilde M \partial_\mu R^T\partial_{\mu} R\right]\]
with 
\[\tilde M=(\tr M)\openone-2M\]
%
Thus we have
\begin{equation}\label{L_alt}
\cL_{\mathrm{spin}}=\frac{1}{2}\tr\left[\tilde I \partial_t R^T\partial_t R-\tilde g \partial_i R^T\partial_i R\right]
\end{equation}
By expressing the matrix elements of $R$ in terms of an orthonormal triad $R_{ab}=\left(\vect{e}_b\right)_a$, with $\vect{e}_a\cdot \vect{e}_b=\delta_{ab}$ this may be written
\begin{equation}\label{L_alt2}
\cL_{\mathrm{spin}}=\frac{1}{2}\sum_{a=1}^3\left[\tilde I_a\left(\partial_t \vect{e}_a\right)^2-\tilde g_a\left(\nabla \vect{e}_a\right)^2\right]
\end{equation}
Recall that for the spin 1 polar phase and for the special value $\eta=0$ in the spin 2 rectangular phase the metric tensor has one zero eigenvalue and two equal non-zero eigenvalues. As a result both $\tilde g$ and $\tilde I$ have \emph{two} vanishing eigenvalues, and Eq.~(\ref{L_alt2}) reduces to the usual $O(3)/O(2)$ sigma model.

\subsection{Equations of motion and spin wave spectrum}

We find the equations of motion corresponding to Eq.~(\ref{L_exp_l_elim}) by writing the variation
\[\left(R^T\delta R\right)_{ab}=-\psi_c\epsilon_{abc}\]
This gives
\begin{eqnarray}
\delta\bm{\omega}_{\mu}&=&\partial_\mu\bm{\psi}-\bm{\psi}\times \bm{\omega}_\mu
\end{eqnarray}
Substitution into Eq.~(\ref{L_exp_l_elim}) leads to the equations of motion.
\begin{equation}\label{eom}
\partial_t \left(I\bm{\omega}_t\right)-\partial_i \left(g\bm{\omega}_i\right)+\bm{\omega}_t\times\left(I\bm{\omega}_t\right)-\bm{\omega}_i\times\left(g\bm{\omega}_i\right)=0
\end{equation}
It follows from their definition that $\bm{\omega}_\mu$ satisfy the (Maurer-Cartan) equation
\begin{equation}\label{mc_eqn}
\partial_\mu\bm{\omega}_{\nu}-\partial_\nu\bm{\omega}_{\mu}+\frac{1}{2}\bm{\omega}_{\mu}\times\bm{\omega}_{\nu}=0
\end{equation}
Note that if $\bm{\omega}_\mu$ is interpreted as a non-abelian gauge field, the above condition corresponds to vanishing field strength, and to the absence of topological defects.

The equations of motion can be linearized by ignoring the right hand side of Eq.~(\ref{mc_eqn}), allowing us to write $\bm{\omega}_\mu=\partial_\mu \bm{\psi}$. The linear equations of motion following from Eq.~(\ref{eom}) are then
\[\partial_t^2 I\bm{\psi}-\partial_i\partial_ig\bm{\psi}=0,\]
a wave equation describing the propagation of three spin wave modes with velocities
\[v_a=\sqrt{\frac{g_a}{I_a}},\qquad a=1,2,3\]
For the spin 2 case Eq.~(\ref{spin2_inertia}) gives
\begin{eqnarray}\label{spin2_inertia}
v_a=
\begin{cases}
\sqrt{2c_1} & \text{tetrahedral phase} \\
\sqrt{g_ac_1-\frac{c_2}{5}} & \text{rectangular phase} 
\end{cases}
\end{eqnarray}
with $g_a$ given by the diagonal elements of Eq.~(\ref{disphenoid_metric}) with $\chi=0$. For the square case ($\eta=(2n+1)\pi/3$) we have $g=\mathrm{diag}\left(4,1,1\right)$ and
\begin{eqnarray}
v_1&=&\sqrt{4c_1-\frac{c_2}{5}}\\
v_2&=&v_3=\sqrt{c_1-\frac{c_2}{5}}
\end{eqnarray}
These results check with Ref.~\onlinecite{ueda2002}, which also includes the normal phonon mode as well as the mode associated with variations of the $\eta$ and $\chi$ parameters in Eq.~(\ref{disphenoid}). In the case of the rectangular phase this latter mode appears gapless in mean field theory, but as explained in Section~\ref{sec:spin2phases} this is the result of an accidental degeneracy that does not persist in the next order of approximation.


\section{Discussion}

We have achieved our goal of providing a framework in which the parameters entering the low energy spin Lagrangian of an arbitrary ordered state of a spinor condensate (with $\phdop_0\vect{S}^{(s)}\phop_0=0$) may be easily calculated. Though we focused on the spin 2 states, any other state can be treated by the same method once the problem of minimizing the mean field energy is solved. The extension of the present formalism to spin ordered Mott insulating phases in which the phase variables are quantum disordered does not present any particular difficulties. 

Perhaps the most interesting problem that we have not addressed in detail relates to the character of topological defects in these systems. The occurrence of nonabelian stabilizer subgroups means that vortices have very novel characteristics~\cite{makela2003,semenoff2007}. We mention one consequence of our work for the \emph{quantum} description of such vortices. The phase factors associated with elements of the stabilizer subgroups that were discussed in Section~\ref{sec:global} will appear in the path integral when vortices are present, as may be seen from Eq.~(\ref{conj_param}). Consider an imaginary time path integral with fields obeying the boundary condition
\[\phi(\br,\tau+\beta)=\phi(\br,\tau)\]
If, as we go from $\tau\to \tau+\beta$, the field at a point $\br$ is subject to a rotation that evolves from $R\to R\gamma$, for $\gamma\in\Gamma$, the $\theta$ variable must increase $\theta\to \theta-\Lambda(\gamma)$ in order to ensure periodicity of the fields, leading to a phase factor $e^{i\rho\Lambda(\gamma)}$ in the path integral ($\rho$ is the density). Ref.~\onlinecite{grover2007} discusses the effect of these phases for the simplest case of the spin 1 polar phase (or rather the Mott insulating phase based upon it), where they are $\pm 1$ and the defects are abelian. The nonabelian case remains unexplored.

\section{Acknowledgments}

The support of the NSF under grant DMR-0846788 is gratefully acknowledged, as well as several useful conversations with Ryan Barnett and Gil Refael.

\appendix

\section{Metric on the order parameter manifold for $\langle \vect{S}\rangle\neq 0$}\label{sec:gauge}

In the general case the distance between states on the ground state manifold has the form Eq.~(\ref{distance}). As mentioned in the text, there is an arbitrariness in the apportioning of phase between the two factors of Eq.~(\ref{gen_param}) that belies a natural gauge structure. In other words, Eq.~(\ref{gen_param}) is unchanged under
\begin{eqnarray*}\label{gauge_trans}
D^{(s)}(R)&\to& D^{(s)}(R)e^{-i\Lambda(R)}\\
\theta&\to& \theta+\Lambda(R)
\end{eqnarray*}
for some arbitrary function $\Lambda(R)$. Let us define a (Berry) vector potential
\begin{eqnarray}\label{berry}
a&=&-i\phdop_0 D^\dagger dD\phop_0\nonumber\\
&=&-\bm\omega\cdot \langle\vect{S}^{(s)}\rangle_0
\end{eqnarray}
where $\langle \cdots \rangle_0$ denotes an expectation in the state $\phi_0$, $\omega_a=\frac{1}{2}\epsilon_{abc}\left(R^T d R^{\vphantom{T}}\right)_{cb}$, and $d$ denotes the exterior derivative (we find it convenient to use the language of differential forms). The vector potential allows us to define a covariant derivative $\mathfrak{d}_a\equiv d-ia$. The metric tensor $d\phdop\otimes d\phop$ can then be cast in the form
\begin{widetext}
\begin{eqnarray}\label{gen_metric}
d\phdop\otimes d\phop&=&
\phdop_0  d  D^\dagger \otimes d D\phop_0 +d\theta\otimes d\theta\nonumber+d\theta\otimes a+a\otimes d\theta \nonumber \\
&=&-\phdop_0  D^\dagger \mathfrak{d}_a  D \otimes D^\dagger \mathfrak{d}^{\vphantom{*}}_a D\phop_0+(d\theta+a)\otimes(d\theta+a)  \nonumber  \\
&=&\omega_a\otimes \omega_b \mathfrak{g}_{ab}+(d\theta+a)\otimes(d\theta+a) 
\end{eqnarray}
%
The  two terms in Eq.~(\ref{gen_metric}) are manifestly gauge invariant under the transformation Eq.~(\ref{gauge_trans}). The gauge invariant metric tensor $\mathfrak{g}_{ab}$ that appears in the first term takes the form~\cite{provost1980}
\begin{equation}\label{gauge_metric}
\mathfrak{g}_{ab}=\frac{1}{2}\langle\{S^{(s)}_a-\langle S^{(s)}_a\rangle_0,S^{(s)}_b-\langle S^{(s)}_b\rangle_0\}\rangle_0.
\end{equation}
In the second term, $a$ by itself is gauge dependent, but the associated field strength is not
\begin{eqnarray}\label{field_strength}
da&=&-d\bm\omega\cdot\langle \vect{S}^{(s)}\rangle_0\nonumber\\
&=&\frac{1}{2}\epsilon_{abc}\omega_a\wedge\omega_b\langle S^{(s)}_c\rangle_0
\end{eqnarray}
where in the second step we have used the Maurer-Cartan equation Eq.~(\ref{mc_eqn}). Eq.~(\ref{field_strength}) has a more familiar form, as may be seen by introducing a unit vector $\vect{m}_0$ parallel to $\langle S^{(s)}\rangle_0$. Then we have
%
\begin{eqnarray}
\epsilon_{abc}\omega_a\wedge\omega_b m_{0,c}&=&\epsilon_{\alpha\beta\gamma}\left(\omega_a\epsilon_{a \alpha\alpha'}\right)m_{0,\alpha'}\wedge \left(\omega_b \epsilon_{b \beta\beta'}\right)m_{0,\beta'}m_{0,\gamma}\nonumber\\&=&\epsilon_{\alpha\beta\gamma}\left(R^TdR\right)_{\alpha\alpha'}m_{0,\alpha'}\wedge\left(R^TdR\right)_{\beta\beta'}m_{0,\beta'}m_{0,\gamma}\nonumber\\
&=&\epsilon_{\alpha\beta\gamma}\left(R^TdR\right)_{\alpha\alpha'}m_{0,\alpha'} \wedge\left(R^TdR\right)_{\beta\beta'}m_{0,\beta'}\left(R^TR\right)_{\gamma\gamma'}m_{0,\gamma'}\nonumber\\
&=&\epsilon_{\alpha\beta\gamma}dR_{\alpha\alpha'}m_{0,\alpha'} \wedge dR_{\beta\beta'}m_{0,\beta'}R_{\gamma\gamma'}m_{0,\gamma'}\nonumber\\
&=&\epsilon_{\alpha\beta\gamma}dm_\alpha \wedge dm_\beta m_\gamma
\end{eqnarray}
\end{widetext}
where $\vect{m}=R\vect{m}_0$, and in the penultimate line we have used the fact that the determinant of the rotation matrices is unity. As a result
\begin{equation}\label{gen_mh}
da=\frac{1}{2}\epsilon_{abc}\langle S^{(s)}_a\rangle_0 dm_b \wedge dm_c,
\end{equation}
which generalizes the Mermin-Ho relation in Eq.~(\ref{v_fix}) to an arbitrary spin state.

An important example is provided by the ferromagnet, for which the state $\phop_0$ is a fully spin polarized (coherent) state, and we have
\[g=\frac{s}{2}\left(\delta_{ab}-m_{0,a}m_{0,b}\right)
.\]
Then the first term of Eq.~(\ref{gen_metric}) takes the form
\[\omega_a\otimes \omega_b \mathfrak{g}_{ab}=\frac{s}{2}d\vect{m}\otimes d\vect{m}.\]
The resulting metric sets the form of the Hamiltonian in Ref.~\onlinecite{Lamacraft2008}.



\end{document}